# Electrical Programmable Multi-Level Non-volatile Photonic Random-Access Memory

Jiawei Meng[1], Yaliang Gui[1], Behrouz Movahhed Nouri[2], Gelu Comanescu[2], Xiaoxuan Ma[1], Yifei Zhang[3], Cosmin-Constantin Popescu[3], Myungkoo Kang[4], Mario Miscuglio[1], Nicola Peserico[1] Kathleen A. Richardson[4], Juejun Hu[3], Hamed Dalir[2], Volker J. Sorger[1,2]*

[1]Department of Electrical and Computer Engineering, George Washington University, Washington, DC 20052, USA
[2]Optelligence LLC, 10703 Marlboro Pike, Upper Marlboro, 20772, MD, USA
[3]Department of Materials Science & Engineering, Massachusetts Institute of Technology, Cambridge, MA 02139, USA
[4]CREOL, The College of Optics & Photonics, University of Central Florida, Orlando, FL 32816, USA

*Corresponding Author: sorger@gwu.edu

**Photonic Random-Access Memories (P-RAM) are an essential component for the on-chip non-von Neumann photonic computing by eliminating optoelectronic conversion losses in data links. Emerging Phase Change Materials (PCMs) have been showed multilevel memory capability, but demonstrations still yield relatively high optical loss and require cumbersome WRITE-ERASE approaches increasing power consumption and system package challenges. Here we demonstrate a multi-state electrically-programmed low-loss non-volatile photonic memory based on a broadband transparent phase change material ($Ge_2Sb_2Se_5$, GSSe) with ultra-low absorption in the amorphous state. A zero-static-power and electrically-programmed multi-bit P-RAM is demonstrated on a silicon-on-insulator platform, featuring efficient amplitude modulation up to 0.2 dB/μm and an ultra-low insertion loss of total 0.12 dB for a 4-bit memory showing a 100x improved signal to loss ratio compared to other phase-change-materials based photonic memories. We further optimize the positioning of dual micro-heaters validating performance tradeoffs. Experimentally we demonstrate a half-a-million cyclability test showcasing the robust approach of this material and device. Low-loss photonic retention-of-state adds a key feature for photonic functional and programmable circuits impacting many applications including neural networks, LiDAR, and sensors for example.**

## Introduction

Photonic computing is one of the main answers to the novel and exponentially increasing data processing for artificial intelligence and machine learning[1,2]. While the benefits given by the intrinsic electromagnetic nature of the optical signals as an energy-efficient way to transmit information are clear, those can potentially be hindered by the optoelectrical and electro-optical transductions, as well as by the repeated access to digital and non-volatile memory[3]. This last aspect impacts the overall operation speed while producing considerable additional energy loss[4]. Since performing Neural Networks (NN) inference, such as classification

Machine Learning task, accounts for more than 90% of the computing effort, having weight bank elements that do not require additional energy can reduce the energy consumption of those tasks[5,6]. For these reasons, having a heterogeneously integrated optimized photonic memory, which retains information in a non-volatile fashion, poses a great advantage, especially when implementing NN-performing inference where the trained weights are only rarely updated (i.e., depending on the application daily, monthly, yearly, if ever)[7].

For photonic computing, photonic memories are one of the most important and yet difficult to realize essential devices compatible with Photonic Integrate Circuits (PICs). Previous studies based on photonic crystals, micro-ring, or other actively tuned electro-optic modulators cannot achieve the feature of non-volatility[8–10], which is the key to low-cost, long-term stable photonic memory. Phase-Change Material (PCM) based photonic memories have appeared recently as a competitive candidate for real non-volatile photonic memory[11–14]. PCMs can be switched between two structural states, the amorphous and crystalline states, with distinct optical and electrical properties. Those states can be reversibly cycled under appropriate thermal or optical stimulation with long-term stability[15]. One of the commonly used PCM materials for photonic memory is GST (Ge-Sb-Te) which exhibits a relatively large contrast of both refractive index ($\Delta n$) and optical loss ($\Delta k$) when switching between amorphous and crystalline states.[16,17 18,19] However, GST is characterized by a high absorption coefficient even in the amorphous state of the PCM while P-RAM is set as ERASING state which is much higher than amorphous state absorption of GSSe The passive insertion loss of GSSe based P-RAM will be much lower than GST based P-RAM. Therefore, for large photonic networks such as the one implementing deep Neuro-Networks the multi-layer design requires the photonic memory containing the kernel memory to be very low loss[20], an aspect that cannot be met by GST-based photonic memory due to its high passive absorption coefficient.

In this work, we develop and demonstrate a non-volatile electrically controlled photonic memory based on the phase-change material Ge2Sb2Se5 (GSSe). The memory is electrically programmed by micro-heaters while the stored information is retained in the solid-state domain (i.e. crystallinity of the PCM), whereas the READ operation is optical by passing a signal through the waveguide cladded by the PCM.de The information is stored in the solid state domain, namely a change in the crystallinity (i.e. amorphous and crystalline). Key is the material selection GSSe over other PCM options keeping amplitude-modulation requirements in mind, demanding a high extinction ratio (ER) and low insertion loss (IL). GSSe was selected since it has one of the lowest known optical losses (amorphous state) of PCMs while offering a high ER through the broadband region for telecommunication wavelengths. The amorphous state GSSe is characterized by a remarkably low adsorption coefficient $2.0 \times 10^{-5}$ at 1550 nm wavelength, enabling near-lossless devices monolithically co-integrated with PICs (Fig. 1e). This low absorption coefficient is over two orders of magnitude lower than GST at 1550 nm[21]. Meanwhile, when in its crystalline state, the absorption coefficient increases to 0.14, which results in a high absorption contrast between the two states (below we also discuss multi-bit memory operation of a single GSSe memory element).

Noteworthy, we demonstrate the optical absorption in the amorphous state is vanishingly close to zero when heterogeneously integrated into silicon photonics. Moreover, the relatively low variation of the absorption coefficient changes in each state makes it a promising material for very stable high order multistate devices, avoiding the utilization of high input laser power and extremely low noise equivalent power detectors. Assuming a continuous film, for the fundamental TM mode of the waveguide, the phase transition produces a variation of the effective absorption coefficient equal to 0.015 which corresponds to 0.2 dB/$\mu m$. Meanwhile, all electrical pulsed programming methods through micro-metal heaters for our proposed P-RAM gain a significant advantage in the ease of control compared to all-optical laser heating for PCM writing and resetting. Also from the packaging perspective, electrical control is still one of the best options, especially for future mass implementation of P-RAM in large-scale photonic tensor computing circuits.

**Results**

To demonstrate this photonic non-volatile memory, a thin 40nm thick layer of GSSe film is directly deposited through the thermal evaporator on the top of a planarized silicon waveguide (Fig. 1a). The obtained memory states are programmed by selectively 'WRITE/ERASE' operations of portions of the GSSe film via local electro-thermal heating. Unlike previous approaches using optical control beams[22], here industry-standard micro heaters are deployed by placing multi-layer metal stips near the waveguide with varying the horizontal distance. This allows optimizing ER vs. IL such as by preventing scattering introduced by metal (see Methods, Fig. 1). It is the key for keeping the overall insertion low and allows electrically driven change of GSSe's structural state (crystalline/amorphous) and consequently results in the strong imaginary-part variation of the effective refractive index, leading to a significant optical absorption change.

In this design, heat is applied to the material externally via joule heating of a tungsten-titanium (W/Ti) metal layer in contact with a 20 nm aluminum oxide dielectric layer over the GSSe film to protest GSSe from oxidization, whose mode and thermal profile are simulated (Fig. 1f.) According to the type of transition wanted, different pulse train profiles are applied to the metal wire via electrical connections to the device. With the 3D mode simulation through Comsol (Supplementary Note 7), we optimize the position of the heaters regarding the waveguide to minimize the ohmic losses due to the presence of metal and concurrently lower the threshold voltage for delivering the necessary amount of heat for inducing the phase transition. The optimized heaters configuration consists of two non-plasmonic tungsten resistive heaters placed in contact with a thin spacer of aluminum oxide deposited on top of the GSSe film(Fig. 1g,h). The heaters are placed 500 nm away on the side of the waveguide, thus providing heat to the film locally, which not only lowers the switching threshold but also temporally stores the heat for successive pulses.

To reach the multi-state power output response, a sequence of paired heaters is placed along the waveguide in series. Each pair of heaters are individually tuned to Joule heat the GSSe material locally for solid-state phase transition. Whereas, in the crystalline state, the GSSe becomes much lossier with a linear absorption coefficient of ~0.2 dB/µm obtained by experimental data and close to zero insertion loss in the amorphous state. This configuration takes advantage of this near-lossless characteristic of the GSSe material in its amorphous state, as the optical signal loss in the waveguide is minimal even for long strips. We precisely control the state of each portion of material by tuning each pair of heaters to obtain a stepwise extinction ratio. When N pairs of heaters are placed, a total of N+1 memory states of power intensity response are realized.

Our photonic memories comprise one single 30-nm thin GSSe pad with paralleled pairs of W-Ti micro-heaters arranged along the waveguide, as each pair of heaters correspond to a quantized state (Fig. 1a,b). The double-sided heater design leads to the highest thermal energy efficiency for the phase transition of GSSe. Furthermore, this design prevents the extra optical insertion loss introduced by the metal heaters, since the metal strips are not directly deposited over the waveguide, but instead have a few hundred nm horizontal distance from the side of the waveguide.

We also present an alternative layout (Fig. 1d), that comprises 30-nm-thin and 5000-nm-wide programmable PCM wires arranged in a grating fashion (duty cycle 50%), with a series of single-sided heaters to Joule-heat each PCM wire, exploiting the same electrical local Joule-heating concept. The single-side heater concept shown here is mainly used for optical memory with a high order of bit number resolution which requires a larger amount of GSSe material cells along with the required number of micro-heaters, metal pads, and routing.

By using this layout, an all-electrical controlled 4-bit photonic memory element is implemented. Considering the highest state as the condition in which all GSSe wires are in the amorphous phase, 15 reprogrammable wires are sufficient for implementing the 4-bit memory within an overall length of just 80 $\mu m$ GSSe pad along with the silicon structure, excluding electrical circuitry. The insertion loss, defined as the optical power loss when all the wires are in an amorphous state, is only 0.12 dB for the 4-bit multilevel memory. The optical power transmitted decreases when the GSSe wires are written/SET (switching to crystalline), leading to discrete power levels for each quantized state. The insertion losses for the multistate memory device with different quantization states are measured (Fig. 2a) which realized 16 quantization states for 4bit. The photonic memory implemented in this configuration provides a uniform quantization. For a 4-bit photonic memory, the quantization step is 0.75 dB/state with the total maximum extinction ratio of 12 dB achieved along with 16 output power states.

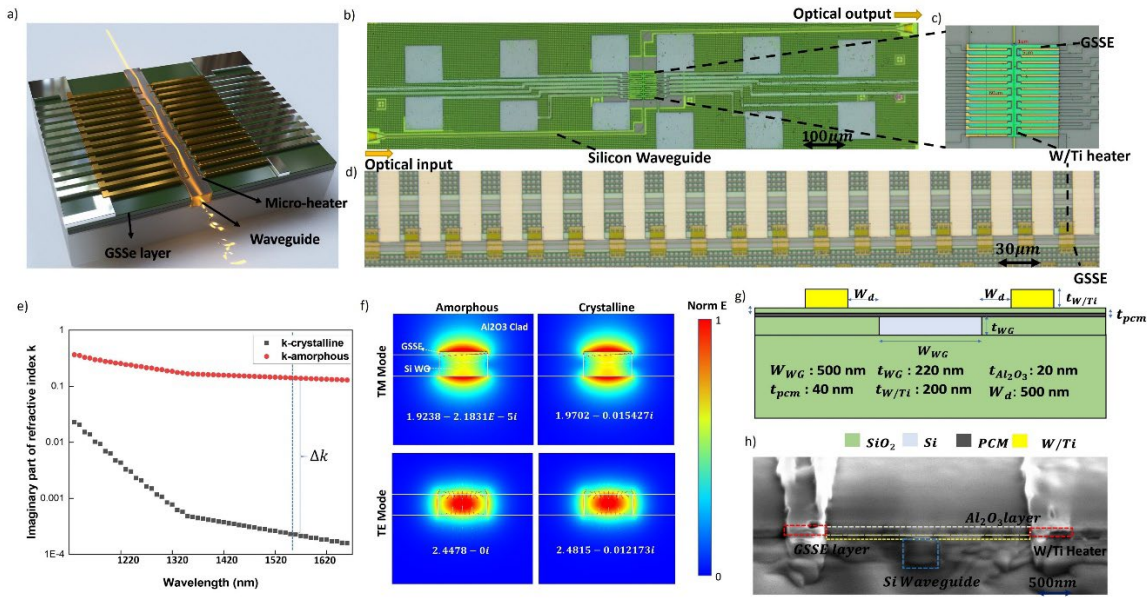

**Figure 1.** Low-loss multi-bit electrically-driven photonic random access memory (P-RAM) on chip. (a) 3D schematic of a planarized waveguide with a 30 nm GSSe layer on top of the waveguide and multiple parallel double-sided tungsten-titanium micro heaters. (b) Detailed optical image of GSSe on waveguide with discrete double-sided heaters (c) Zoom-in detailed image of in b. Discrete double-sided heaters are arranged along the waveguide over the GSSe film. (d) Detailed optical image of GSSe strip array with single-sided heaters for measurement of high order bit memory. (e) Experimentally obtained (ellipsometry) optical properties of GSSe film. Absorption coefficient contrast (imaginary part of the refractive indices of amorphous and crystalline alloys) of GSSe for crystalline and amorphous states at 1550 nm. The GSSe shows a strong unity Δk, while simultaneously showing a small, induced loss at the amorphous state. (f) Normalized electric field

mode profile of hybrid Si-GSSe waveguide for TE and TM mode with 0.54dB/um absorption coefficient between amorphous and crystalline state. The effective refractive index of k in the amorphous state is $-2.18 \times 10^{-5}$ which leads to an exceedingly small unit of passive absorption of the memory. (g) 2D cross-section schematic of the lateral thermoelectric switching configuration to optimize the heater resistance for max heating efficiency with minimum optical scattering. (h) Cross-section SEM image of the device.

To enhance the speed and energy efficiency of electro-thermally switched PCM devices, we optimize the micro-heaters position. We tested different distances between waveguide and heaters, from 0.125 $\mu m$ to 5 $\mu m$ with the previously shown design of double-sided heaters. With the same amount of electrical energy applied to each heater pair, the total extinction ratio achieved decreased with the increasing distance, while concurrently the unit insertion loss introduced by the GSSe cell decreased (Fig. 2b). To balance the phase transition energy efficiency and insertion loss (IL), we calculated the Figure-Of-Merit (FOM) as ER/IL for each distance which is a well-used metric for the evaluation of electro-optical modulator performance[21], and the optimized position (Fig. 2c), indicating 500 nm distance as the best value which is the smallest distance between waveguide and micro-heaters for which metal won't introduce extra optical loss towards the waveguide while remaining high thermal efficiency. At this distance, we compared the FOM of our proposed device along with three other PCM photonic memories (Fig. 2e,f). With the same theoretical unit insertion loss, our devices achieved the highest unit extinction ratio.[23]

To evaluate the endurance of our device, a cyclability measurement is conducted and a total of half-million Writing-Resetting cycles was successfully achieved (Fig. 2d), with stable power responses in either state. The main limitation which prevents memory from achieving higher Writing-Resetting cycles was the failure of micro-heaters on the two sides of the waveguide. With the large number of heating-cool down cycles for GSSe Writing-Resetting, the initial tungsten micro-heaters were easily broken due to oxidization under the fast temperature change (Fig. 2d). To overcome the issue, we replace the heater material from tungsten to tungsten-titanium with a 200nm thick dual layer of aluminum deposited over the W/Ti on the routing part. The Al layer reduces the electrical resistance, enhances heat uniformity, and protects the W/Ti heaters. Meanwhile, a 600 nm thick layer of $Al_2O_3$ is deposited over the device to prevent further oxidation and physical bend of metal.[24] Such structure allows the heater to survive after half-million cycles (Fig. 2b). This is the longest cycle test on PCM integrated into a photonic circuit with stable writing-resetting photonic responses. Despite having demonstrated the record-high cyclability herein, we believe that this microheater-driven P-RAM technology has to potential for yet significantly higher cyclability; since device failure of our devices scales with passivation oxide thickness and quality, further improvements of the cladding will increase cyclability. Formerly reported failure mechanisms of either void formation or elemental segregation may be other factors that should be explored further[25].

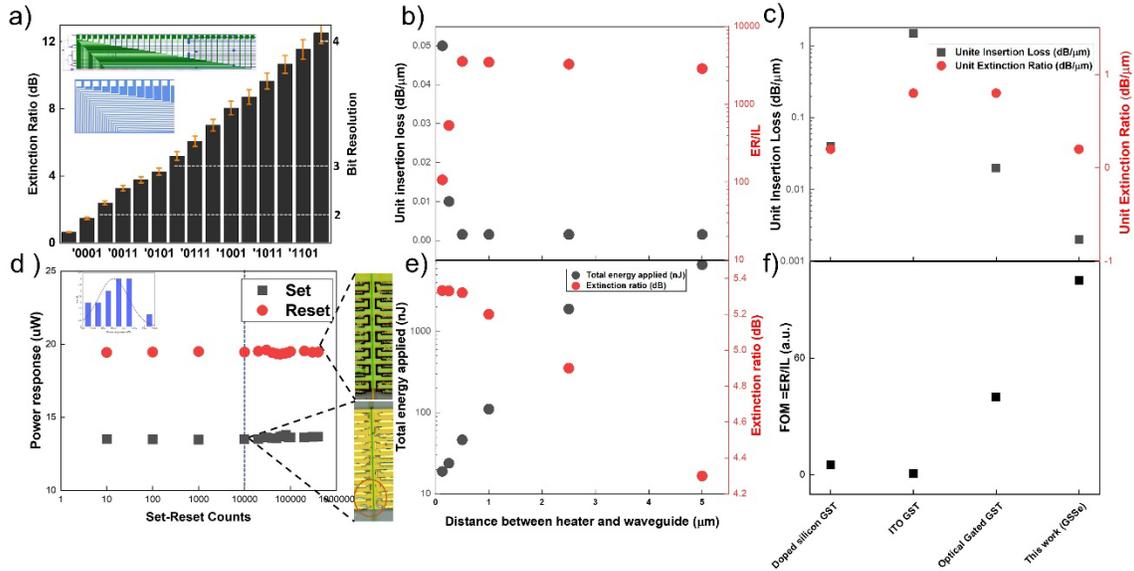

**Figure 2**. P-RAM performance for bit resolution, energy, cyclability, and FOM. (a) Optical power response for a 4-bit photonic memory as a function of digital states, for an increasing number of crystalline-wire the Extinction Ratio (ER) increases linearly and uniformly. (b) Unit insertion loss (IL) and extinction ratio (ER) per unit insertion loss vs. heater position. (c) Unit insertion loss and unit extinction ratio comparison between PCM-based photonic memories. (d) Bi-State optical responses change exceeding 500,000 switching cycles. For heaters exposed to air with no $Al_2O_3$ layer protection, the maximum Writing-Resetting cycles achieved is 10,000 and then heaters were broken due to heavy oxidization or physical deformation as shown. With a thick $Al_2O_3$ layer on top of the heaters, the maximum cycle reached is 500,000 and heaters are still alive as shown in the lower right subfigure. (e) Heater performance vs position of the heater. The distance between the edge of the waveguide and the double heater is swept from 125 nm to 5000 nm. Left axis: Total energy applied vs. heater-waveguide distance for reaching 6-dB extinction ratio. Right axis: With the same applied energy, the ER change corresponds to the heater position. (f) Figure of merit comparison for different PIC-based non-volatile photonic memories. Doped silicon GST[26,27], ITO GST[27], Optical Gated GST[22]

For GSSe material, the amorphization temperature is the melting point (>900K), while for crystallization a certain temperature (~600K) must be applied and kept constant for approximately 20$\mu s$.[28] Crystallization is achieved by applying the pulse setting (Fig. 3c) to keep the material temperature consistent in the desired range for over 20$\mu s$, while the amorphization is achieved by adding a threshold voltage 10-12V (~2 $\mu s$) to the local heater up to 900K. The voltage range takes into consideration the fabrication variability of micro-heaters. The real-time continuous Writing-Resetting measurement for two states of memory (Fig. 3a).

Since the total extinction ratio that we want to achieve is proportional to the area of the GSSe cell covering the waveguide, the phase transition time required is also proportional to the desired extinction ratio, for the different thermal volumes of PCM material. We then experimentally map the amount of ER that can be achieved as a function of transition time at the falling edge of the real-time normalized optical power transmission trace (Fig. 3a), and the achieved effective extinction ratio compare to the delay time after the last setting pulse (Fig. 3b). To achieve 0.2 dB ER, 0.5 ms is needed compared to 500 ms required for a 6 dB total ER response. After the setting pulses are applied, the delay time needed for reaching different extinction ratio (Fig. 3b) indicates a non-linear relationship. The memory working speed is determined by the desired extinction ratio since the thermal expansion takes time and the area of PCM phase transition determines the amount of extinction ratio introduced by the phase transition.

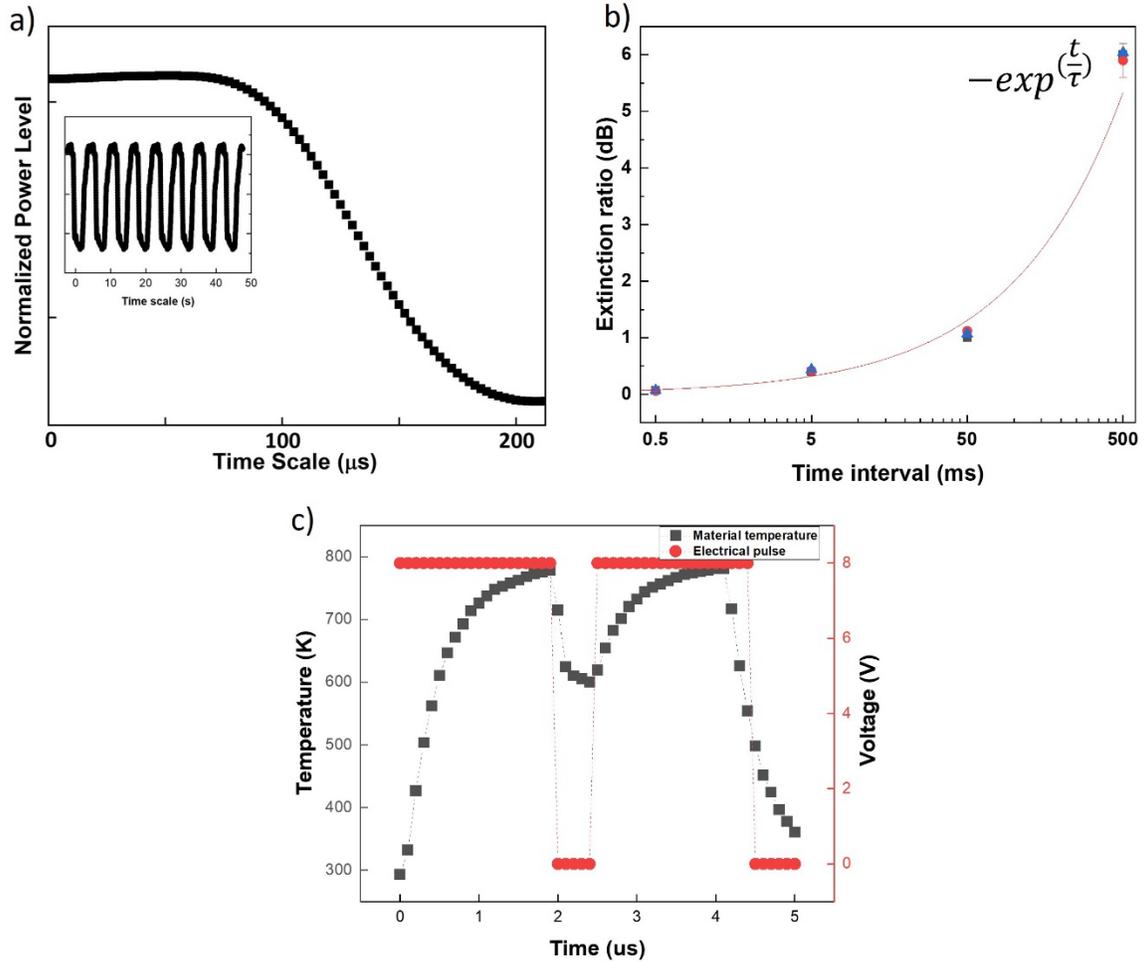

**Figure 3.** P-RAM speed response and writing pulse set up. (a) Time-dependent trace of normalized optical power transmission with the power response down edge in μs level. (b) The time taken for reaching different levels of extinction ratio is varied as shown in the figure from 0.5 ms to 500 ms. (c) Simulated pre-programmed voltage pulses are applied for each heater and a two-sided neighbor works simultaneously for GSSe to transient from amorphous to crystalline state.

Besides the concept of multi-heater pairs that we proposed for the multi-states optical memory, here we propose another design concept for the optical memory (Fig. 4a). As we described, the multi-state optical power response is realized by tuning the ratio of GSSe film over the waveguide through local Joule-heating to introduce a different level of insertion loss[29]. Based on this theory, we developed the non-equal heater pair memory cell. The shorter heater on the right side works for the amorphization of the GSSe cell. The COMSOL electrical-thermal simulation results (Fig. 4b), indicate that with different energies applied to the heater, the above-melting-point hot area changes, which introduces the different amorphous areas and so the different absorption levels. The fitting equation (Fig. 4b) indicates the different extinction ratio achieved is propagated to the thermal decay time $\frac{1}{e}\tau$ as long as the device heat structure gives us a hint about how the real terminal structure of P-RAM influences the rime response of extinction ratio achieved. The numeric results (Fig. 4d) display more clearly the distribution of hot area (>900K) with the increasing of electrical energy applied to the micro-heater. On the other side, the longer heater on the left side works as the resetting button to change the full GSSe cell into its crystalline state and erases all the previously stored information set by the right heater. The experimental result for 6 different states (2.58-bit) has been achieved (Fig. 4c) for different energy levels from 80 to 400 nJ and further measurement and optimization is will keep going on for the higher-order and goal to the 5-bit 32 states and the theoretical highest bit resolution is mainly limited by the resolution of photodetector for which can distinguish small neighbor extinction ratio over the noise level.

The main advantage of this design is the reduction of electrical tracks and pads, allowing for an improved integration into photonic integrated circuits such as photonic neural network (NN), or any other optical on-chip designs which need the local optical data storage without extra E-O-E conversion. Within a single PCM material and a single pair of set-reset heaters, a multi-states response can be realized which significantly helps to reduce the footprint for the implementation of large bit number memory. The setting and resetting energy could be saved simultaneously since the total area of PCM that need to be heated up is smaller than the parallel memory cell structure. Meanwhile, only one pair of heaters is needed for this design which reduces the footprint of metal routing, and the complexity of the electrical package for the PIC chip will be significantly reduced as well as the fabrication cost. On the other hand, the main drawback is the difficulty of programming compared to the multi-heater pair design. In the first designs, the programming pulse setting for each heater pair is fixed, and the ultimate power response is the result of an accumulation for all parallel paired heaters. For this design, there is not an *a priori* known relationship between ER and applied energy. To get the multi-state power response with a fixed step size, a pulse setting for each state needs to be individually set, introducing an extra program difficulty.

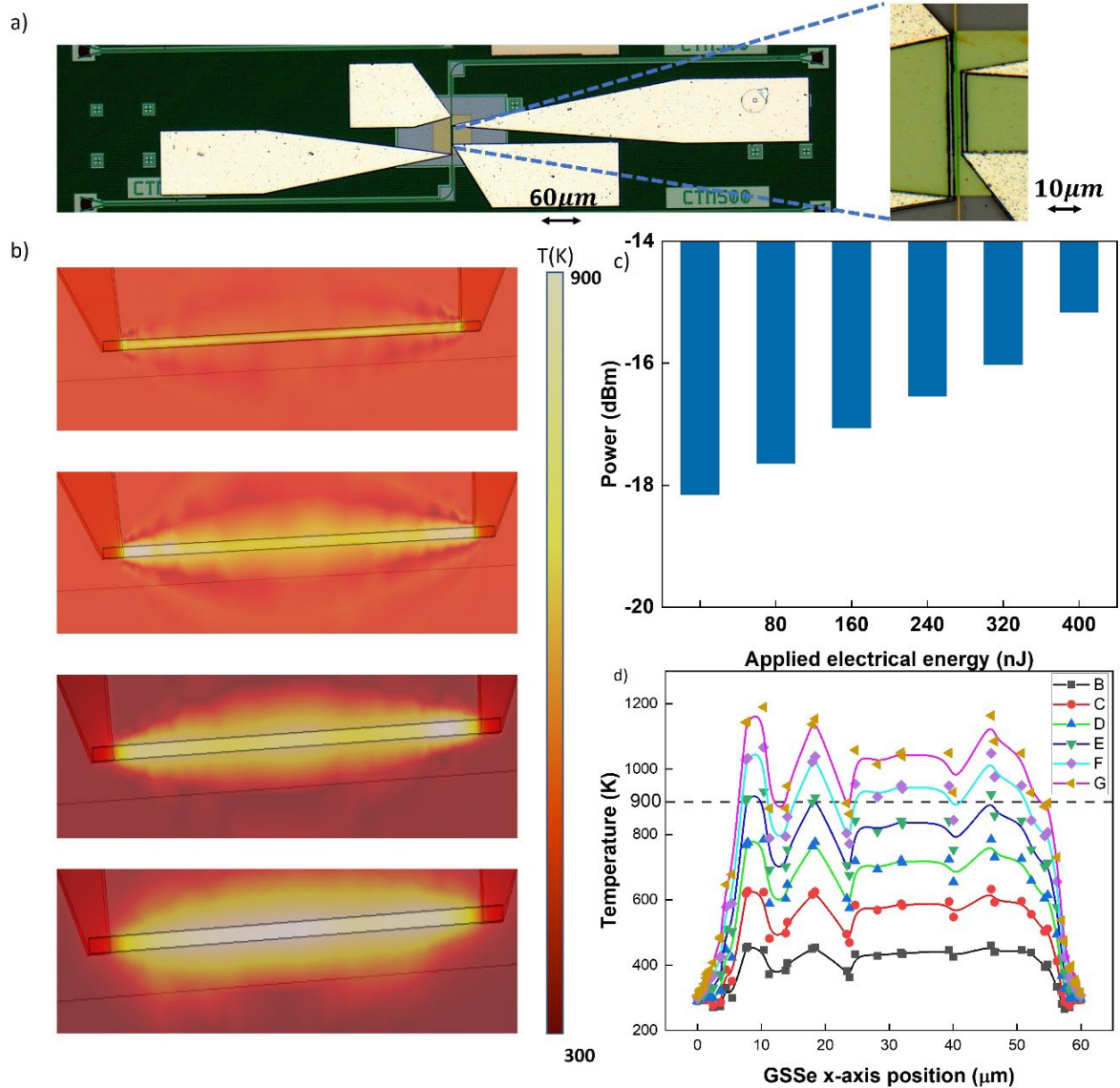

**Figure 4**. Single heater pair, Multi-states power response P-RAM. (a) Schematic of single pair heater multi-level power response device with a different crystalline-amorphous ratio on a single GSSe pad. Detailed optical image of asymmetric paired heaters stands along the waveguide over the GSSe pad. The longer heater on the left side is used as the resetting heater to change the phase of all GSSe film from amorphous to crystalline. The shorter heater on the right side is used as the setting heater to control the area of amorphous state film by applying different levels of Joule-heating energy to the W-Ti microheater. As more energy is applied, a larger area of GSSe film will be transient to the amorphous state and the total absorption will decrease. Based on different levels of absorption, a multi-level power response function is achieved. (b) COMSOL thermal simulation of temperature distribution along with the heater also indicates the area ratio of GSSe material in the amorphous and crystalline state. (c) Measured 6-states power responses achieved. (d) GSSe material temperature distribution along with the heater from COMSOL thermal simulation with different energy applied to micro-heater.

**Discussion**

In this paper, we present a novel ultralow insertion loss phase change material GSSe, implementing non-volatile electrical-controlled (k)-only modulation photonic memory as various reconfigurable devices following a similar concept. Compared to MZI or micro-ring based index (n)-only modulation schematic, this amplitude (k)-only schematic based device has more stable operation, for example as the temperature of the chip varies. From the electrooptic modulators' well-known tradeoff when deploying resonators, namely that ER improves with the finesse however becoming spectrally narrow and requiring stabilization control circuitry raising complexity, similarly apply PCM-based P-RAMs as well. Furthermore, we prove the device's endurance with over half a million switching cycles. The number is mainly limited by the durability of metal heaters, as they were physically broken after a large number of heating and cooling cycles. Following that, we improved the lifetime of heaters with a thick oxide layer covered on the top. Novel structures and devices could be optimized to enhance further the photonic memory cyclability by improving the design of the materials stack for the heaters.

A few key P-RAM performance characteristics have been compared with two other demonstrated P-RAM approaches, as shown in Table 1. Though this work has larger setting energy and smaller unit extinction ratio compared to the all-optical setting GST-based P-RAM, we hold the best figure-of-merit (ER/IL) due to our ultra-low insertion loss benefitting from the transparent GSSe material and a novel double-sided metal heater design. Moreover, we have successfully demonstrated half-million Writing-Resetting cycles with very stable performance which is far more than other P-RAM's cyclability results, as shown in Table 1.

*Table 1*. Main P-RAM performance comparison. IL = Insertion loss; ER = Extinction ratio of signal modulation. FOM figure-of-merit

| Material | Programming Method | to Energy ($nJ/dB$) | ER ($dB/\mu m$) | Unit IL ($dB/\mu m$) | Performance (i.e. ER/IL) | Implementation Complexity | Writing-Resetting Cycles |
|---|---|---|---|---|---|---|---|
| GST [22] | Optical absorption | 1.0 | 0.8 | 0.02 | 40 | High | 5000 |
| GST [26,30] | Doped silicon heater | 7.6 | 0.2 | 0.04 | 13 | Medium | 3800 |
| GSSe (This work) | On-chip integrated heater | 1.5 | 0.2 | <0.002 | 70 | Low | ~500,000 |

From simulations, we expecte an extinction ratio of 0.4 $dB/\mu m$, while from the experimental demonstration we obtaine about 0.2 dB/µm. The main reason for this difference is due to the heat distribution applied to the PCM cell through the micro-heater, as not the whole PCM reached the transition temperature. The heat spread follows an ellipse shape which results in a non-uniform temperature map (Fig. 4b), that caused a lower extinction ratio compared to the simulations. The different crystalline-amorphous ratios caused by the non-uniform heating lead to our second proposed P-RAM design, which compromises a smaller volume of PCM, and by so a more uniform heat distribution.

As we described before, the total extinction ratio of our P-RAM could be achieved is based on the length of GSSe cell which could be transitioned through a micro-heater. Then the highest bit resolution that could be achieved by each device is limited by the minimum detected dynamic extinction ratio for every single state through an optical power meter. Based on our current measurement setup, the minimum detectable power range is 35 pW which means that we could achieve 1 binary state as small as over 35 pW difference in theory. Then for traditional 4- or 5-bit memory, the length of the active region for each memory could be sub-micrometer long and could be cascaded for different bit resolutions required.

The non-volatility of our P-RAM results in zero static power consumption for state maintenance and exceptionally low insertion loss introduced by active PCM material GSSe. Meanwhile, the setting energy of our P-RAM is also relatively low, computed around 1.5 $nJ/dB$. As we discussed previously, the bit resolution is limited by the minimum dynamic extinction ratio in dB that could be detected over the system noise level, which means that the required energy for each bit Writing-Resetting and the required footprint could be as small as nW level as shown in Table 1 for our device.

In large-scale photonic computing architectures, such as high order matrix MAC operation required for deep neural networks, the stringent energy requirements motivate the implementation of multiple photonic memories for weight bank[5,31]. For these challenges, our devices can perform even orders of magnitude better than volatile memories in terms of energy consumption and footprint.

Besides the low operating energy consumption for high dimension photonic tensor operations, our proposed P-RAM takes advantage of all-electrical micro-heaters, and by so reducing the packaging complexity compared to all-optical laser heating P-RAMs compared in Table 1. When tens of thousands of P-RAMs need to be implemented, electrical control is the only feasible way for memory programming and large-scale photonic circuit packaging[32].

**Conclusion**
In summary, we have experimentally demonstrated a new class of electrically driven optical non-volatile memory with near-zero insertion loss and low power consumption, which exploits the unique optical properties of the phase-change material GSSe to achieve zero-static-power consumption and low-dynamic-power consumption in ultra-compact devices. Two different P-RAM designs with similar basic concepts were demonstrated which could be utilized in low-energy programmable photonic integrated circuits. Our results are promising for applications in photonic computing architectures such as weight banks in optical neural networks, optical switching for telecommunication, quantum networks, and others.

**Methods**

Device fabrication
20 nm GSSe thin film layer was deposited by using single-source thermal evaporation and a 20 nm layer of $Al_2O_3$ was deposited by using atomic layer deposition (ALD) as a protective coating to prevent GSSe from oxidation. The tungsten-titanium microheater was fabricated in the nanofabrication and imaging center at George Washington University. A 200 nm think tungsten titanium layer is sputtered. Then another 200 nm think Al is deposited over the W/Ti route to decrease the overall resistance, increase the heat spread over the micro-heaters, and to protect the W/Ti layer from oxidation. Then a thick 400 nm $Al_2O_3$ layer is deposited over the full circuit using the ALD for oxidation prevention. Contact pad windows are opened using oxide layer plasma dry etch for electrical probes to connect with circuits for micro-heaters driving.

Electro-thermal simulation/optical mode simulation and microheater modeling:
The Joule heating process and heat dissipation model were performed using a three-dimensional finite-element method in COMSOL Multiphysics. We used the AC/DC Joule-heating module coupled with the heat transfer module, which accounts for surface radiation as well as thermal boundary resistance.

**Acknowledge**

This work was performed in part at the George Washington University Nanofabrication and Imaging Center (GWNIC). Thin film material analysis is supported from NIST Center for Nanoscale Science and Nanotechnology (CNST), and J.A. Woollam Co.
V.J.S. is supported by AFOSR (FA9550-20-1-0193) under the Presidential Early Career Award in Science and Engineering (PECASE).